\documentstyle[sprocl]{article}
\def\lsim{\mathrel{\rlap{\lower4pt\hbox{\hskip1pt$\sim$}}
    \raise1pt\hbox{$<$}}}         %less than or approx. symbol
\def\gsim{\mathrel{\rlap{\lower4pt\hbox{\hskip1pt$\sim$}}
    \raise1pt\hbox{$>$}}}         %greater than or approx. symbol

\input psfig.sty

\def\Journal#1#2#3#4{{#1} {\bf #2}, #3 (#4)}

\def\NPA{{\em Nucl. Phys.} A}
\def\NP{\em Nucl. Phys.}

\def\PRL{\em Phys. Rev. Lett.}

\def\PRC{{\em Phys. Rev.} C}

\def\PR{\em Phys. Rep.}
\def\JPG{\em J. Phys. G}
\def\JPA{\em J. Phys. A}

\def\be{\begin{equation}}
\def\ee{\end{equation}}
\def\bea{\begin{eqnarray}}
\def\eea{\end{eqnarray}}
\begin{document}

\title{THE CANONICAL NUCLEAR MANY-BODY PROBLEM AS A RIGOROUS EFFECTIVE THEORY}

\author{W. C. HAXTON AND C.-L. SONG}

\address{Institute for Nuclear Theory, Box 351550, and
Department of Physics, Box 351560\\
University of Washington, Seattle, WA 98195, USA\\
E-mail: haxton@phys.washington.edu}

%%%%%%%%%%%%%%%%%%%%%%%%%%%%%%%%%%%%%%%%%%%%%%%%%%%%%%%%%%%%%%
% You may repeat \author \address as often as necessary      %
%%%%%%%%%%%%%%%%%%%%%%%%%%%%%%%%%%%%%%%%%%%%%%%%%%%%%%%%%%%%%%

\maketitle\abstracts{ The shell model is the standard tool for
addressing the canonical nuclear many-body problem of nonrelativistic
nucleons interacting through a static potential.   
We discuss several of the uncontrolled approximations that are
made in this model to motivate a different approach, one based
on an exact solution of the Bloch-Horowitz
equation.  We argue that the necessary self-consistent 
solutions of this equation can be obtained efficiently by a
Green's function expansion based on the Lanczos algorithm.
The resulting effective theory is carried out for the 
simplest nuclei, d and ${}^3$He, using realistic NN interactions
such as the Argonne $v18$ and Reid93 potentials, in order to contrast
the results with the shell model.  We discuss the wave function
normalization, the evolution of the wave function as the
``shell model" space is varied, and the magnetic elastic
effective operator.  The numerical results show a simple 
renormalization group behavior that differs from effective 
field theory treatments of the two- and three-body problems.
The likely origin of this scaling is discussed.}

\section{Introduction}
  
In many text books the shell model (SM) is motivated by the analogy
with Brueckner's treatment of nuclear matter.  While the exact
many-body Hamiltonian operates in an infinite Hilbert space
\begin{equation}
H = {1 \over 2} \sum_{i,j=1}^A (T_{ij} + V_{ij}),
\end{equation}
where $T_{ij}$ is the relative nonrelativistic kinetic energy
operator and $V_{ij}$ the nucleon-nucleon potential, the 
SM Hamiltonian acts in a restricted space and 
employs a softer ``effective" potential,
\begin{equation}
H_{SM} = {1 \over 2} \sum_{i,j=1}^A (T_{ij} + V^{eff}_{ij}).
\end{equation}
Motivating $H_{SM}$ is the notion that the determination of
$V^{eff}$ might be simpler than solving the original A-body problem:
the foundation of Brueckner theory is that high-momentum contributions
to the wave function might be integrated out in a rapidly converging
series in $\rho_{nuclear}$ or, equivalently, in the number of 
nucleons in high-momentum states interacting at one time outside
the SM space.

The SM thus represents explicitly $\sim$ 60\% of the 
wave function that resides at long-wavelengths, treating the 
A-body correlations important to collective modes by
direct diagonalization.  Implicitly the high-momentum components 
are swept into a rather poorly defined ``effective interaction,"
often determined empirically.  The strength
of the SM resides in the first of these two aspects: the technology developed
for direct diagonalizations is quite remarkable, including 
recent progress in Lanczos-based methods \cite{poves},
in treatments of light nuclei involving many shells \cite{barrett},
and in Monte Carlo sampling \cite{koonin,otsuka}.  Its 
weakness is the numerous uncontrolled approximations that 
become apparent when one tries to view the shell model as a 
faithful effective theory (ET).  The thesis of this talk is that
the same numerical strides that have advanced shell model
diagonalizations now allow us to remove these uncontrolled
approximations.  The resulting ET has many
differences with the shell model and many similarities to
the effective field theories under discussion at this
workshop.

Among the SM uncontrolled approximations are the following:\\
1) Even in lowest order, where only the pairwise interaction 
of high-momentum nucleons is included in $H^{eff}$, the 
functional form of the resulting effective interaction is 
not as simple as assumed in the SM,
\begin{equation}
\langle | H^{eff} | \rangle_{SM} \equiv \langle \alpha \beta | H^{eff} | \gamma \delta \rangle
\end{equation}
where the Greek symbols label single-particle shell-model states.
For example, if the Slater determinants are formed from 
harmonic oscillator states, the two-body matrix elements must
carry an additional index labelling the total number of quanta
in the configuration on which $H^{eff}$ operates \cite{zheng}.
Thus $H^{eff}$ reduces to the shell-model form only
when that index is unnecessary, e.g., when a lowest-order calculation is 
restricted to a single shell.  Beyond lowest order,
three-, four-, and higher-body operators are successively added
to $H^{eff}$.\\
2) Typically $H^{eff}$ lacks the symmetries of the original bare
$H$, e.g., translational invariance and Hermiticity (though the
latter is often enforced by hand). \\
3) SM wave functions are orthogonal and normed to unity.
In ET the effective wave functions are naturally defined as the
restrictions of the true wave functions $|\Psi_i \rangle$ to the
model space
\begin{equation}
|\Psi_i \rangle \begin{array}[t]{c} \longrightarrow \\
ET \end{array} \langle SM | \Psi_i \rangle  | SM \rangle \equiv |\Psi_i^{eff} \rangle.
\end{equation}
Thus the norms are less than unity and orthogonality, which holds
for the true wave functions, is lost when these wave functions
are restricted to the model space.\\
4) Shell model interactions frequently depend on fictitious 
parameters such as ``starting energies," introduced to 
adjust the energy denominator in the two-body G-matrix or 
to account for intermediate-state average energies when the
two-body G-matrix is iterated to produce some approximation
to a higher order $H^{eff}$. \\
5) Perhaps most serious, the important issue of effective 
operators is almost never addressed in a meaningful way.  In
many cases practitioners adopted a phenomenological $H^{eff}$
which, while successful in producing spectra, provides no
diagrammatic basis for calculating effective operators or
wave function normalizations.  Even in cases where 
$H^{eff}$ is derived from some underlying NN interaction, 
the practice is generally to then employ bare operators.
In some well-studied cases, such as allowed $\beta$ decay in
the $1p$ and $2s1d$ shells, it is then recognized that a 
phenomenological renormalization (e.g., $g_A \rightarrow$ 1)
of operators greatly improves agreement with experiment.
But the origin of this renormalization and its evolution with
momentum transfer $q$ are left unclear.  The situation is 
very unsatisfactory and undercuts the shell model as a 
predictive tool. \\

\section{Self-consistent Bloch-Horowitz Solutions}
  
We consider the cononical nuclear structure problem of 
nonrelativistic point nucleons interacting through a realistic
NN interaction, such as the Argonne $v18$ \cite{wiringa} and Reid93 \cite{reid93}
potentials.  The question is whether the uncontrolled approximations
in the shell model can be removed, leaving a more complicated
but still tractable effective theory.  The approach involves three
major steps:\\

\noindent
$\bullet$ Formulating a treatment of effective interactions 
and operators that exploits the basic assumption in Brueckner
theory --- that interactions at high momenta can be integrated
out in a cluster expansion (essentially an expansion in 
$\rho a^3$, where $\rho$ is the nuclear density and $a$ an
interaction range) --- but is otherwise exact.  The convergence
of the expansion could then be tested numerically and should
depend on the operator under study and the momentum transfer.
The goal would be to distinguish fully converged results 
from those which require higher order calculations.\\

\noindent
$\bullet$ To find numerical tricks for implementing this
formulation, demonstrating their validity in cases (e.g,
A=2,3,4) where the expansion can be carried to all orders,
so that the answers should then agree with Faddeev and
other exact methods.\\

\noindent
$\bullet$ To imbed the formulation in a heavier nucleus, 
where the cluster expansion can be carried out only partially.\\

There is some reason for optimism that if the first two goals
can be achieved, the third might yield very accurate results: the Argonne group 
cluster variational Monte Carlo effort on $^{16}$O appeared to
yield nearly exact results when clusters up to A = 5 were
included.  In this talk our efforts on the first two points
will be described.  In particular, we will be able to 
contrast an exact ET of the deuteron and ${}^3$He with the
shell model to illustrate the shortcomings of the later:
we think the differences are surprising.

The approach is sketched in Fig. 1.  The Hilbert space is
divided into a long-wavelength ``shell model'' space, defined
by some energy scale $\Lambda_{SM}$, and a high-momentum 
space.  One can truncate the latter at some scale $\Lambda_\infty
\sim 3$ GeV, characteristic of the cores of realistic 
potentials, as above this energy, excitations make a negligible
contribution.  That is,
\begin{equation}
H^{eff}_{(i)}(\Lambda_{SM},\Lambda_{\infty}) \begin{array}[t]{c}
\longrightarrow \\ \Lambda_\infty \mathrm{~large} \end{array}
H^{eff}_{(i)}(\Lambda_{SM}).
\end{equation}
All correlations within the ``SM'' space are included, but the
high-momentum correlations in the excluded space are limited 
to n-body, where n is the cluster size.  Thus the lowest order
effective interaction is
\begin{equation}
H^{eff}_{(n=2)} \equiv H^{eff}_{(n=2,0)}
\end{equation}
It corresponds to embedding the A-body ladder diagram of Fig. 1b
between SM states: A-2 of the nucleons are spectators, with
the remaining pair scattering via a two-body ladder.
The notation (n=2,0) states that the
two-body cluster has no explicit dependence on the nuclear 
density, varying as $\rho^0$.
The n=3 A-body ladder of Fig. 1c is similarly
\begin{equation}
H^{eff}_{(n=3)} \equiv H^{eff}_{(n=2,1)} + H^{eff}_{(n=3,0)}
\end{equation}
where $H^{eff}(n=2,1)$ is the two-body part of the three-body
ladder
\begin{equation}
\langle \alpha \beta | H^{eff}_{(n=2,1)} | \alpha' \beta' \rangle =
\sum_{\gamma \leq k_F} \langle \alpha \beta \gamma | H^{eff}_{(n=3)}
| \alpha' \beta' \gamma - \alpha' \gamma \beta' + ... \rangle,
\end{equation}
where $k_F$ denotes the Fermi level. 
This decomposition -- which is done only to emphasize the content
of the cluster expansion -- illustrates that $H^{eff}_{(n=3)}$
contains $H^{eff}_{(n=2,0)}$ as well as a correction to the two-body interaction
that depends linearly on the density and is obtained by identifying one
ingoing SM single-particle leg of the three-body ladder with one outgoing leg,
summed over all occupied states.  It also contains a true
three-body piece $H^{eff}_{(n=3,0)}$, where three SM ingoing single-particle states 
connect to three distinct outgoing states 
after undergoing a series of scatterings outside the SM
space.  The point is simple pedagogy: treatments of successively
larger clusters in the high-momentum space correct the lowest-order
two-body $H^{eff}_{(n=2,0)}$ by adding terms proportional to the
$\rho$, $\rho^2$, etc., in the spirit of Brueckner theory.  It
also adds true three-body terms, true four-body terms, etc.
Thus an expansion through four-body clusters yields
$H^{eff}_{(n=2,2)}$, the two-body interaction corrected through order $\rho^2$,
$H^{eff}_{(n=3,1)}$, the three-body interaction through order $\rho$, and $H^{eff}_{(n=4,0)}$,
a density-independent true four-body interaction.

\begin{figure}[htb!]
\psfig{bbllx=-3.0cm,bblly=3.0cm,bburx=14cm,bbury=22cm,figure=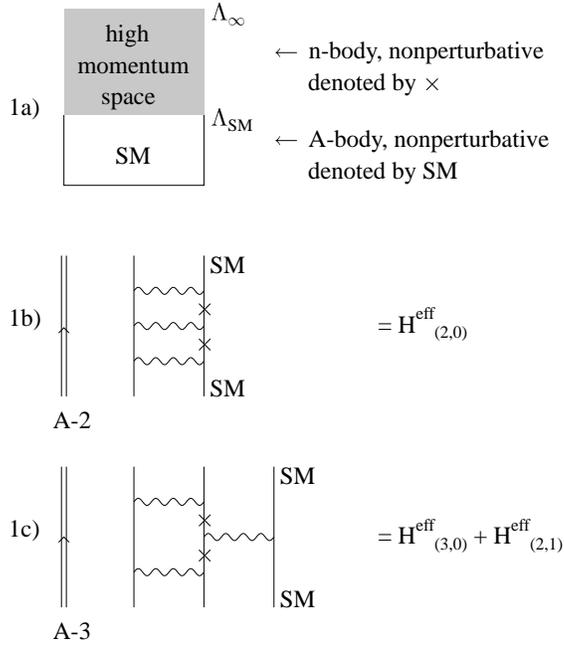,height=3.5in}
\caption{Cluster expansion of the effective interaction.}
\end{figure}
  
The calculation begins with a definition of the ``SM" space.  The
goals of handling bound states and of generating an effective 
interaction that is translationally invariant leaves one sensible
choice, many-body states constructed from harmonic oscillator
Slater determinants.  To exploit the relative/center-of-mass separability
of harmonic oscillator Slater determinants, one must separate the
SM and high-momentum spaces so that all configurations satisfying
\begin{equation}
E \leq \Lambda_{SM} \hbar \omega
\end{equation}
are retained in the former.  For example, a SM calculation of $^{16}$O with
$\Lambda_{SM} = 4 + \Lambda_0$, where $\Lambda_0$ is the number of
quanta in the $^{16}$O closed shell, would include all
$4 \hbar \omega$ configurations, e.g.,
$0p0h$, $2p2h$, and $4p4h$ excitations of nucleons from the $1p$ shell into the
$2s1d$ shell, $1p1h$ excitations of a $1s$ shell nucleon into 
the $3s2d1g$ shell, etc.  One can define the projection operator
onto the high-momentum space by
\begin{equation}
Q_{SM} = Q(\Lambda_{SM},b).
\end{equation}
where $b$ is the oscillator parameter.  Thus the included or ``SM"
space is defined by two parameters, $\Lambda_{SM}$ and $b$.  The preservation of 
translational invariance is also important numerically, as it
reduces the two-body ladder to an effective one-body problem, etc.

The resulting Bloch-Horowitz equation \cite{BH} is then
\begin{eqnarray}
H^{eff} &=& H + H {1 \over E - Q_{SM}H} Q_{SM} H \nonumber \\
H^{eff} |\Psi_{SM} \rangle &=& E |\Psi_{SM} \rangle~~~|\Psi_{SM} \rangle
 = (1-Q_{SM}) |\Psi \rangle
\end{eqnarray}
where $|\Psi \rangle$ is the exact wave function and $H |\Psi \rangle
= E |\Psi \rangle$.  The difficulty posed by this equation is the
appearance of the unknown energy eigenvalue in the equation for
$H^{eff}$.  Thus this system must be solved self-consistently.
Note that there is no explicit reference to the harmonic oscillator
in this equation: it enters only implicitly through $Q_{SM}$ in
distinguishing the long-wavelength ``SM" space from the remainder
of the Hilbert space.

There is an extensive literature on this and similar equations,
often involving a division of $H$ into an unperturbed $H_0$ and
a perturbation $H_1 = H-H_0$ \cite{bk,ko}.  There are well-known
pathologies with this division involving the effects of near-by
intruder states on the perturbation expansion \cite{sw}.  
Here we explore another approach that is nonperturbative and 
involves, in effect, a computer summation of diagrams.  The
method is based on the Lanczos algorithm and offers a remarkably
simple solution to the issue of self-consistency.

In the Lanczos algorithm a basis for representing a Hamiltonian
is formed recursively in such a way that the resulting 
Hamiltonian is tridiagonal.  Given a Hermitian operator $H$ and
and an initial normalized vector $|v_1\rangle$, the successive steps are
\begin{eqnarray}
H|v_1\rangle &=& \alpha_1 |v_1\rangle + \beta_1 |v_2\rangle \nonumber \\
H|v_2\rangle &=& \beta_1 |v_1\rangle + \alpha_2 |v_2\rangle
+ \beta_2 |v_3\rangle \nonumber \\
H|v_3\rangle &=& ~~~~~~~~~~~\beta_2 |v_2\rangle + \alpha_3 |v_3\rangle
+ \beta_3 |v_4\rangle~~etc. 
\end{eqnarray}
so that the $H$ takes the form
\begin{equation}
H \rightarrow \left( \begin{array}{cccc} \alpha_1 & \beta_1
& 0 & \cdots \\ \beta_1 & \alpha_2 & \beta_2 & \cdots \\
0 & \beta_2 & \alpha_3 & \cdots \\ \vdots & \vdots & \vdots & 
\end{array} \right) \left( \begin{array}{c} |v_1\rangle \\
|v_2\rangle \\ |v_3\rangle \\ \vdots \end{array} \right)
\end{equation}
The remarkable property of this algorithm has to do with 
truncating the process in Eq. (12)
after $n$ steps, where $n$ can be much smaller than the dimension
of the Hilbert space.  The resulting truncated matrix in Eq. (13)
then contains the information needed to reconstruct the exact
2$n$-1 lowest moments of H over the eigenspectrum.  As extremum
eigenvalues are crucial to higher moments, one common application
of the Lanczos algorithm is in determining such eigenvalues
and their associated eigenfunctions.  Another is to begin with
the vector $|v_1\rangle = \hat{O}|g.s.\rangle$ and then use the
algorithm to calculate the moments of the response of the ground
state $|g.s.\rangle$ to the operator $\hat{O}$.  A small number
of moments, e.g., $\sim$ 100, often is sufficient to construct
a response function with a numerical resolution comparable to that achieved
experimentally.

A third application \cite{haddock} is in constructing fully
interacting Green's functions.  One finds
\begin{equation}
{1 \over E-H} |v_1\rangle = g_1(E) |v_1\rangle + g_2(E) |v_2\rangle
+ \cdots
\end{equation}
where the $g_i(E)$ are continued fractions that depend on 
${\alpha_i,\beta_i}$ and where E appears only as a parameter.
For example,
\begin{equation}
g_1(E) = {1 \over {E - \alpha_1 - {\beta_1^2 \over
E - \alpha_2 - {\beta_2^2 \over E - \alpha_3 - \beta_3^2} \cdots}}}
\end{equation}
  
It follows that the Bloch-Horowitz equation can be solved
self-consistently with only a single solution of the effective
interactions problem, even in cases where multiple bound
states are needed.  The procedure is: \\
  
\noindent
$\bullet$ For each relative-coordinate vector in the SM space
$|\gamma\rangle$, form the excluded-space vector
$|v_1\rangle \equiv Q_{SM}H|\gamma\rangle$ and the corresponding
Lanczos matrix for the operator $Q_{SM}H$.  Retaining the
resulting coefficients ${\alpha_i,\beta_i}$ for later use,
construct the Green's function for some initial guess for
$E$ and then the dot product with $\langle \gamma' | H$ to find
$\langle \gamma' | H^{eff}(E) | \gamma \rangle$. \\
  
\noindent
$\bullet$ Perform the ``SM" calculation to find the
desired eigenvalue $E'$ which, in general, will be different
from the guess $E$.  Using the stored ${\alpha_i,\beta_i}$,
recalculate the Green's function for $E'$ and $H^{eff}(E')$
then redo the ``SM" calculation.  Repeat until convergence,
i.e., until the input $E'$ in the Green's function equals 
the desired output ``SM" eigenvalue.\\
  
\noindent
$\bullet$ Then proceed to the next desired bound state, e.g., the
first excited state, and repeat the above step.  Note that it is not
necessary to repeat the $H^{eff}$ calculation.  The eigenvalue
taken from the ``SM" calculation is, of course, that of the
first excited state.  The procedure then generates distinct
$H^{eff}(E')$s for each desired state.\\

The attractiveness of this approach is that the effective 
interactions part of the procedure, which is relatively time consuming
as it requires one to perform a large-basis Lanczos calculation
for each relative-coordinate starting vector in the ``SM" space,
is performed only once.  The diagonalization in
the model space is generally much faster: modern workstations 
can handle even large-dimension shell model calculations
(sparse matrices of $d \sim 10^6$) quickly ($\sim$ 30 minutes).
In practice we found that self-consistency is achieved 
easily: six to eight cycles is typical.  (More cycles
are required for states with small binding energies.)
Thus it is quite practical to derive the exact $H^{eff}(E)$s
for a series of bound states.

Now we discuss the results of applying this procedure to the
simplest nuclei, d and ${}^3$He, carrying the above process to
completion (two-body and three-body ladders, respectively).
The motivation is two-fold: demonstrate the numerical procedures
we described above, and provide for the first time exact effective
theory results that can be compared to those of the shell model.

The harmonic oscillator mode expansion must be sufficient 
to represent both the long-distance tails of bound states and
the short-distance ``hard core" scattering predicted by
realistic NN potentials.  (The Argonne v18 ${}^1S_0$ potentials
are shown in Fig. 2.)  
Inclusion of high-momentum states through $\Lambda_\infty \sim 50$
yields a deuteron binding energy accurate to $\sim$ 60 keV;
extending this to $\sim$ 140 produces a result accurate to
one keV, provided one chooses an oscillator parameter that 
optimizes the convergence.  (However, as we will discuss
in Section 3, there appears to be a simple scaling with
$\Lambda_\infty$ that allows one to extrapolate results to 
$\Lambda_\infty \rightarrow \infty$.  When this is done, the
small binding energy differences that result at $\Lambda_\infty 
= 140$ for a reasonable choices of $b$ all but disappear.
We will return to this point in Section 2.)
Fig. 3 shows the rate of convergence as a function of $b$ and $\Lambda_\infty$.
The rate of convergence for ${}^3$He is similar to that for 
the deuteron: a $\sim$
60 keV energy accuracy is achieved at a $\Lambda_\infty$ of 
$\sim$ 50.

\begin{figure}[htb!]
\psfig{bbllx=0.0cm,bblly=9.0cm,bburx=18cm,bbury=22.5cm,figure=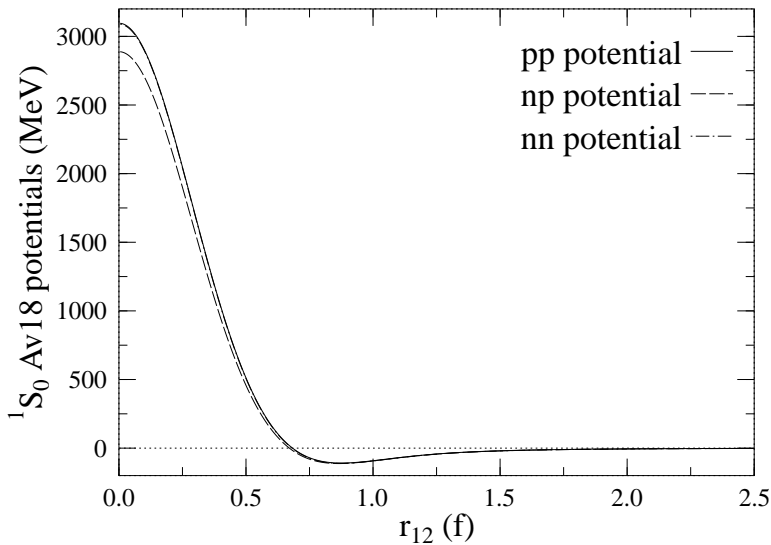,height=3.0in}
\caption{The Argonne $v18$ ${}^1S_0$ potentials.}
\end{figure}

The binding energies and operator matrix elements for simple
systems like ${}^3$He can, of course, be calculated
by other methods, e.g., Faddeev techniques or Green's function
Monte Carlo.  We thus want to stress that the point of the 
following discussion is to do analogous calculations in the
context of an effective theory, so that we begin to see the
shortcomings of conventional techniques like the shell model
as well as possibilities for overcoming those shortcomings.
A first test of the techniques outlined above is to solve the
Bloch-Horowitz equation for some SM-like space to then see
if the resulting self-consistent energy is, indeed, the 
correct value.  For model spaces of 2, 4, 6, and 8$\hbar \omega$
in the case of the deuteron we obtained a binding energy of
-2.224 MeV (using $\sqrt{2} b=1.6f$ and $\Lambda_\infty = 140$).  The exact result is -2.2246 MeV.  Similar agreement
was obtained for ${}^3$He.

\begin{figure}[htb!]
\psfig{bbllx=0.cm,bblly=9.0cm,bburx=18cm,bbury=22.5cm,figure=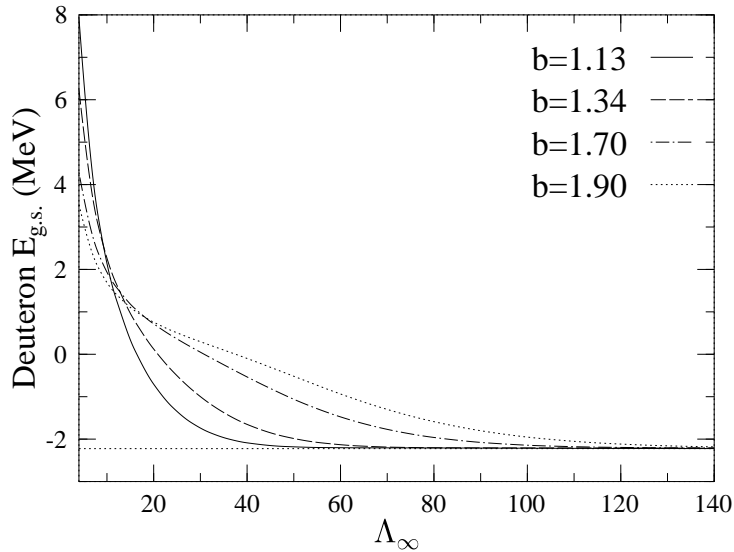,height=3.0in}
\caption{Deuteron ground state energy convergence as a function of
$\Lambda_\infty$ for several choices of the oscillator parameter $b$.
This b is defined as in the independent-particle SM: the corresponding
b for relative motion is $\sqrt{2}$ times the values shown.
It is clear that if an inappropriate size is chosen for the basis states
(e.g., b=1.9), the rate of convergence can be greatly slowed.}
\end{figure}

More interesting is the evolution of the wave functions, 
shown in Tables~\ref{table:one} and ~\ref{table:two}.  Various ET calculations were done in
small model spaces, analogous to shell model spaces, consisting
of all 2$\hbar \omega$ configurations, all 4$\hbar \omega$
configurations, etc.  For each such space we then solved the
Bloch-Horowitz equation via the Lanczos Green's 
function method described above, iterating the shell model
calculation until the self-consistent energy is fully converged.
The deuteron and ${}^3$He calculations involve two- and
three-body ladder sums in the excluded space, yielding 
sets of two- and three-body ``SM" matrix elements of $H^{eff}$
for the model spaces.  The deuteron calculation is rather
trivial; for $\Lambda_\infty \sim 50$ the ${}^3$He BH 
calculation involves a dense matrix of dimension $\sim 10^4$,
still rather modest by current SM standards.  The matrix is
dense because we work in relative Jacobi coordinates, rather
than an m-scheme, utilizing standard Talmi-Brody-Moshinshy 
methods \cite{mosh}.  (See Ref. \cite{song} for details.)
The results in Table~\ref{table:two} were obtained with approximately 100 Lanczos
iterations: it is apparent that the convergence is then quite good.  The wave functions must be 
normalized according to Eq. (11): this involves calculating
unity as an effective operator.  We will return to this point
below.

\begin{table}
\caption{ET results for the deuteron ground state wave function 
calculated with the Argonne $v18$ potential.  The columns on the
right correspond to different choices of the ET model space,
the analog of a SM space.  The rows correspond to the resulting amplitudes
for the designated, selected configurations $\mid n, {}^SL_J \rangle$.  The quantities
within the parentheses are the square of the norm of the 
effective wave function, e.g., the probability that the deuteron
resides in the corresponding ``SM" space.}
\label{table:one}
\vspace{0.2cm}
\begin{center}
\begin{tabular}{|r|r|r|r|r|r|r|}
\hline
\hline
 & \multicolumn{6}{c|}{amplitude} \\ \cline{2-7}
basis state & 0$\hbar \omega$  & 2$\hbar \omega$ &
4$\hbar \omega$ & 6$\hbar \omega$ & 8 $\hbar \omega$ & exact \\ \cline{2-7}
 & (65.9\%) & (79.5\%) & (86.1\%) & (91.3\%) & (93.0\%) & (100\%) \\ \hline
 $\mid$ 1, ${}^3S_1 \rangle$ & 0.81155 & 0.81154 & 0.81155 & 0.81155 & 0.81152 & 0.81155 \\ \hline
 $\mid$ 2, ${}^3S_1 \rangle$ & 0.00000 & -0.31483 & -0.31483 & -0.31483 & -0.31482 & -0.31483 \\ \hline
 $\mid$ 1, ${}^3D_1 \rangle$ & 0.00000 & 0.19524 & 0.19524 & 0.19524 & 0.19523 & 0.19524 \\ \hline
 $\mid$ 3, ${}^3S_1 \rangle$ & 0.00000 & 0.00000 & 0.24945 & 0.24945 & 0.24944 & 0.24945 \\ \hline
 $\mid$ 4, ${}^3S_1 \rangle$ & 0.00000 & 0.00000 & 0.00000 & -0.20851 & -0.20850 & -0.20851 \\ \hline
 $\mid$ 5, ${}^3S_1 \rangle$ & 0.00000 & 0.00000 & 0.00000 & 0.00000 & 0.12596 & 0.12596 \\ \hline\hline
\end{tabular}
\end{center}
\end{table}

The tables show the lovely evolution of the wave function in ET,
an evolution quite unlike that of typical shell model calculations.
The wave functions obtained in different model spaces agree 
over overlapping parts of their Hilbert spaces.  Thus as one
proceeds through 2$\hbar \omega$, 4$\hbar \omega$, 6$\hbar \omega$,
... calculations, the ET wave function evolves only by adding 
new components in the expanded space.  The normalization of the
wave function grows accordingly.  Thus, for ${}^3$He,
the 0$\hbar \omega$ ET calculation contains 0.311 of the full
wave function in the effective space; the 0+2+4$\hbar \omega$
result is 0.700.  

This evolution will not arise in the standard SM because the wave
function normalization is set to unity regardless of the model
space.  It will also not arise for a second reason, illustrated in
Table~\ref{table:three}.  The matrix elements of $H^{eff}$ are crucially 
dependent on the model space: the listed results for ${}^3$He
show that a typical matrix $\langle \alpha | H^{eff} | \beta \rangle$
changes very rapidly under modest expansions of the model space,
e.g., from 2$\hbar \omega$ to 4$\hbar \omega$.  Yet it is 
common practice in the shell model to expand calculations
by simply adding to an existing SM Hamiltonian new interactions
that will mix in additional shells.  
We suspect the behavior found for ${}^3$He is generic in ET
calculations: it arises because a substantial fraction of the
wave function lies near but outside the model space (e.g., see Table~\ref{table:two}).  An
expansion of the model space changes the energy denominators 
for coupling to some of these configurations, and moves 
other nearby configurations from the excluded space to the model
space.  Naively, relative changes in effective interaction matrix
elements of unity are expected.

\begin{table}
\caption{As in Table~\ref{table:one}, only for ${}^3$He.  The basis states are
now designated somewhat schematically as $\mid N, \alpha \rangle$,
where $N$ is the total number of oscillator quanta and 
$\alpha$ is an index representing all other quantum numbers.}
\label{table:two}
\vspace{0.2cm}
\begin{center}
\begin{tabular}{|r|r|r|r|r|r|r|}
\hline
\hline
 & \multicolumn{6}{c|}{amplitude} \\ \cline{2-7}
state & 0$\hbar \omega$ & 2$\hbar \omega$ &
4$\hbar \omega$ & 6$\hbar \omega$ & 8 $\hbar \omega$ & exact \\ \cline{2-7}
 & (31.1\%) & (57.4\%) & (70.0\%) & (79.8\%) & (85.5\%) & (100\%) \\ \hline
 $\mid 0, 1 \rangle$ & 0.55791 & 0.55791 & 0.55791 & 0.55795 & 0.55791 & 0.55793 \\ \hline
 $\mid 2, 1 \rangle$ & 0.00000 & 0.04631 & 0.04613 & 0.04618 & 0.04622 & 0.04631 \\ \hline
 $\mid 2, 2 \rangle$ & 0.00000 & -0.48255 & -0.48237 & -0.48243 & -0.48243 & -0.48257 \\ \hline
 $\mid 2, 3 \rangle$ & 0.00000 & 0.00729 & 0.00731 & 0.00730 & 0.00729 & 0.00729 \\ \hline
 $\mid 2, 4 \rangle$ & 0.00000 & 0.16707 & 0.16698 & 0.16706 & 0.16706 & 0.16708 \\ \hline
 $\mid 2, 5 \rangle$ & 0.00000 & 0.00566 & 0.00564 & 0.00565 & 0.00565 & 0.00566 \\ \hline
 $\mid 2, 6 \rangle$ & 0.00000 & -0.00017 & -0.00017 & -0.00017 & -0.00017 & -0.00017 \\ \hline
 $\mid 4, 1 \rangle$ & 0.00000 & 0.00000 & -0.02040 & -0.02042 & -0.02043 & -0.02047 \\ \hline
 $\mid 4, 2 \rangle$ & 0.00000 & 0.00000 & 0.11267 & 0.11274 & 0.11275 & 0.11289 \\ \hline
 $\mid 4, 3 \rangle$ & 0.00000 & 0.00000 & -0.04191 & -0.04199 & -0.04208 & -0.04228 \\ \hline
 $\mid 4, 4 \rangle$ & 0.00000 & 0.00000 & 0.28967 & 0.28978 & 0.28978 & 0.29001 \\ \hline
 $\mid 4, 5 \rangle$ & 0.00000 & 0.00000 & 0.01059 & 0.01059 & 0.01059 & 0.01059 \\ \hline
 $\mid 4, 6 \rangle$ & 0.00000 & 0.00000 & -0.00213 & -0.00212 & -0.00211 & -0.00210 \\ \hline
 $\mid 4, 7 \rangle$ & 0.00000 & 0.00000 & 0.00998 & 0.01000 & 0.01000 & 0.01000 \\ \hline
 $\mid 4, 8 \rangle$ & 0.00000 & 0.00000 & -0.11319 & -0.11327 & -0.11330 & -0.11335 \\ \hline
 $\mid 4, 9 \rangle$ & 0.00000 & 0.00000 & 0.08446 & 0.08447 & 0.08446 & 0.08448 \\ \hline
 $\mid 4, 10 \rangle$ & 0.00000 & 0.00000 & -0.08613 & -0.08626 & -0.08632 & -0.08638 \\ \hline
 $\mid 4, 11 \rangle$ & 0.00000 & 0.00000 & -0.00210 & -0.00211 & -0.00211 & -0.00211 \\ \hline
 $\mid 4, 12 \rangle$ & 0.00000 & 0.00000 & -0.00252 & -0.00254 & -0.00256 & -0.00257 \\ \hline
 $\mid 4, 13 \rangle$ & 0.00000 & 0.00000 & -0.00020 & -0.00020 & -0.00020 & -0.00020 \\ \hline
 $\mid 4, 14 \rangle$ & 0.00000 & 0.00000 & -0.00010 & -0.00010 & -0.00010 & -0.00010 \\ \hline
 $\mid 4, 15 \rangle$ & 0.00000 & 0.00000 & -0.00012 & -0.00013 & -0.00012 & -0.00012 \\ \hline \hline
\end{tabular}
\end{center}
\end{table}
  
\begin{table}
\caption{Selected BH 3-body effective interaction matrix elements for ${}^3$He,
in MeV, illustrating the strong dependence on the ``SM" space.  The Argonne
v18 potential was used.}
\label{table:three}
\vspace{0.2cm}
\begin{center}
\begin{tabular}{|r|r|r|r|r|}
\hline
\hline
  & 2$\hbar \omega$ &
4$\hbar \omega$ & 6$\hbar \omega$ & 8 $\hbar \omega$  \\ \hline
$\langle 0, 1 \mid H^{eff} \mid 2, 1 \rangle$ & -4.874 & -3.165 & -0.449 & 1.279 \\ \hline
$\langle 0, 1 \mid H^{eff} \mid 2, 5 \rangle$ & -0.897 & -1.590 & -1.893 & -2.208 \\ \hline
$\langle 2, 1 \mid H^{eff} \mid 2, 2 \rangle$ & 6.548 & -2.534 & -4.144 & -5.060 \\ \hline \hline
\end{tabular}
\end{center}
\end{table}

Now we turn to the question of operators.  The standard procedure
in the SM is to calculate nuclear form factors with 
bare operators, or perhaps with bare operators renormalized 
according to effective charges determined phenomenologically
at $q^2$ = 0, using SM wave functions normed to 1.
As we now have a series of exact effective interactions
corresponding to different model spaces, we can test
the validity of this approach.  The results for the elastic
magnetic form factors for the deuteron and $^3$He are shown in
Figs. 4 and 5.  One sees in each case that by the time one 
reaches a momentum transfer $q \sim 2.5/f$, random numbers are
being generated: bare operators used in conjunction with exact
effective wave functions generate results that differ by an
order of magnitude, depending on the choice of the model space.
This is not surprising, of course.  If one considers the 
operation
\begin{equation}
e^{i \vec{q} \cdot \vec{r}} \sigma \tau_3 | g.s. \rangle 
\end{equation}
at momentum transfers $\gsim 2k_F$, where $k_F$ is the Fermi 
momentum, most of the resulting amplitude should reside outside
the long-wavelength model space, in any simple view of the 
nucleus.  That is, the strength resides entirely in the effective
contributions to the operator.  If these components are ignored,
the results have to be in error.

\begin{figure}[htb!]
\psfig{bbllx=0.cm,bblly=9.0cm,bburx=18cm,bbury=22.5cm,figure=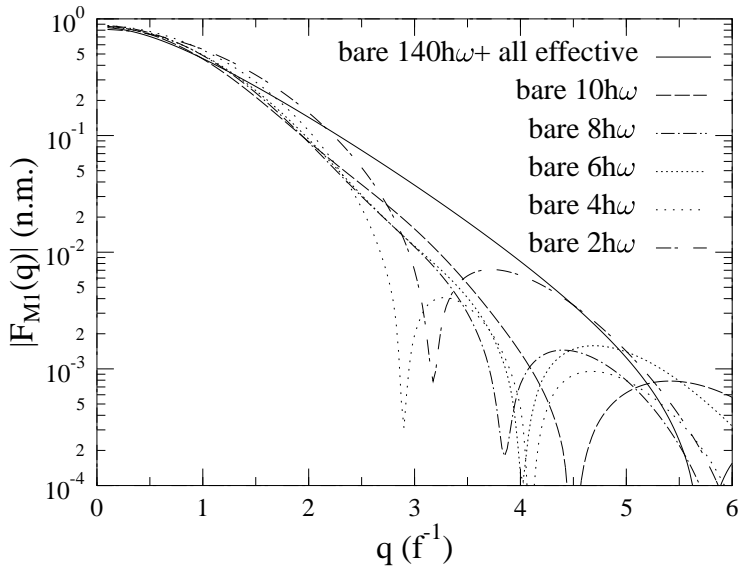,height=3.0in}
\caption{The magnetic elastic form factor for the deuteron
calculated with the exact $H^{eff}$, SM wave functions normalized
to unity, and a bare operator are compared to the exact result
(solid line).  When effective operators and the proper wave
function normalizations are used, all results become identical
to the solid line.}
\end{figure}

Clearly the effective interaction and effective operator have to
be treated consistently and on the same footing.  If $\hat{O}$
is the bare operator, one finds
\begin{equation}
\langle \Psi_f | \hat{O} | \Psi_i \rangle \equiv \langle \Psi_f^{eff}
 | \hat{O}^{eff} | \Psi_i^{eff} \rangle 
\end{equation}
where
\begin{equation} 
\hat{O}^{eff} = (1 + HQ_{SM} {1 \over E_f - HQ_{SM}}) \hat{O}
(1 + {1 \over E_i - Q_{SM}H}Q_{SM}H)
\end{equation}
and where the effective wave function normalization of $|\Psi_i^{eff} \rangle$ and $|\Psi_f^{eff} \rangle$, mentioned
earlier, must be determined using the effective operator $\hat{1}$,
e.g.,
\begin{equation}
1 = \langle \Psi_i | \Psi_i \rangle = \langle \Psi_i^{eff} |
(1 + HQ_{SM}{1 \over E_i - HQ_{SM}})(1 + {1 \over E_i - Q_{SM}H}
Q_{SM}H) | \Psi_i^{eff} \rangle
\end{equation}
These expressions can be evaluated with the Lanczos Green's function
methods described earlier.  When this is done, all of the effective calculations,
regardless of the choice of the model space, yield the same result,
given by the solid lines in Figs. 4 and 5.

\begin{figure}[htb!]
\psfig{bbllx=0.0cm,bblly=9.0cm,bburx=18cm,bbury=22.5cm,figure=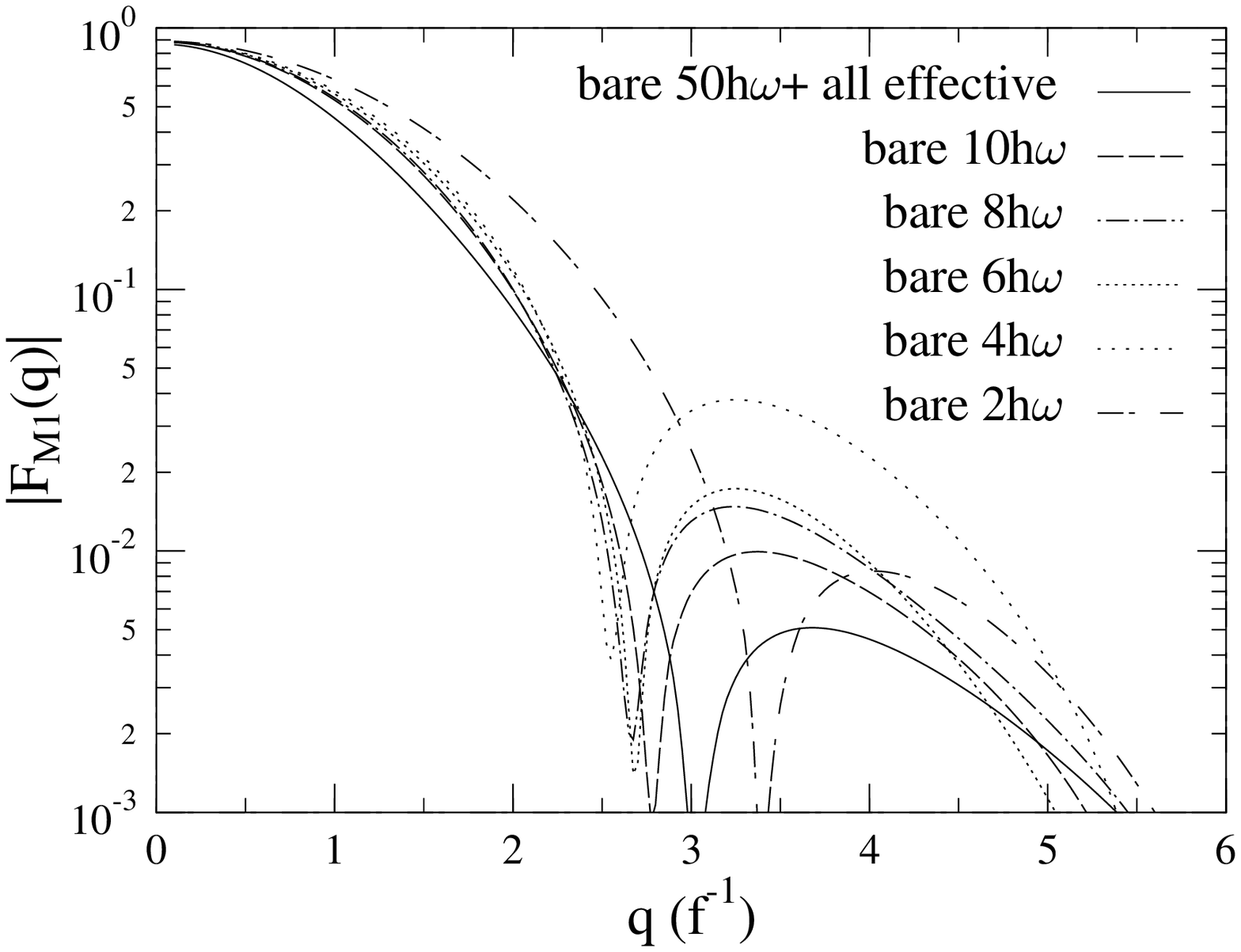,height=3.0in}
\caption{As in Fig. 4, only for ${}^3He$.}
\end{figure}

We would argue, based on this example, that many persistent
problems in nuclear physics --- ranging from the renormalization
of $g_A$ in $\beta$ decay \cite{brown} to the systematic differences
between measured and calculated M1 form factors at Bates \cite{peterson} ---
very likely are due to our naive treatments of operators.  It
should be apparent from the above example that no amount of work
on $H^{eff}$ will help with this problem.  What is necessary is
a diagrammatic basis for generating $H^{eff}$ that can be 
applied in exactly the same way to evolving $\hat{O}^{eff}$.
From this perspective, phenomenological derivations of $H^{eff}$
by fitting binding energies and other static properties of 
nuclei are not terribly helpful, unless one intends to 
simultaneously find phenomenological renormalizations for 
each desired operator in each $q^2$ range of interest.

\section{Numerical Renormalization Group Speculations}

While the Lanczos Green's function method is reasonably elegant,
the calculations described above have a ``brute force" aspect in
that the high-momentum ladders must be summed to very large
$\Lambda_\infty$.  In the of the deuteron we carried the sums
to $\Lambda_\infty = 140$ to assure one keV accuracy.  The 
calculation for ${}^3$He was stopped at $\Lambda_\infty = 52$,
with a resulting binding energy of -6.842 MeV.  As our
calculations are otherwise exact, this is an upper bound:
a variational principle operates as $\Lambda_\infty$ is
increased.  The corresponding Green's function Monte Carlo
result is -6.87 $\pm$ 0.03 MeV; Faddeev results for the v18
potential give -6.895 $\pm$ 0.001 MeV.

The ${}^3$He calculations can (and will) be carried further:
the existing matrices are only of dimension 12000.
However, there are alternatives to the brute-force approach.
One rather simple idea is to evaluate, instead of the three-body
ladder, the difference of the three-body ladder and the 
appropriate sum over the three corresponding two-body ladders,
evolving this difference with $\Lambda_\infty$.  The resulting
$H^{eff}$ then would be obtained by adding to this difference
the two-body ladder.  While this is a tautology, our expectation
is that the difference will converge more quickly than the full
3-body ladder as a function of $\Lambda_\infty$, since the 
difference would start with all pair wave functions being
properly correlated.  If one were able to truncate the 3-body
difference calculation at a lower $\Lambda_\infty$ than the
corresponding 2-body ladders, considerable efficiency would be
gained.  This possibility will soon be studied.

Another possibility has to do with an observation about the 
way both ground-state energies and matrix elements of $H^{eff}$
evolve with $\Lambda_\infty$.  Our calculations are based on the
division of the Hilbert space into two sectors
\begin{equation}
0 \leq \Lambda \leq \Lambda_{SM}~~~~\Lambda_{SM} < \Lambda \leq \Lambda_\infty
\end{equation}
While the deuterium calculation was done with a large value of
$\Lambda_\infty$, we can bring it to a much lower value in order
to study the evolution with $\Lambda_\infty$ (see Fig. 3).
When the numerical results were examined, it was found that they
scaled with $\Lambda_\infty$ simply, approximately as an exponential in 
$\Lambda_\infty^2$.  For example, calculations were
done for the Argonne $v18$ potential and $b$ = 1.7f, with
$\Lambda_\infty$ = 46, 50, and 54, to which we fit
the three-parameter functional form
\begin{equation}
E_{g.s.}(\Lambda_\infty) = -2.175 \mathrm{MeV} + 3.178 \mathrm{MeV} e^{-1.055 (\Lambda_\infty/50)^2}
\end{equation}
Thus -2.175 MeV is identified as the binding energy for 
$\Lambda_\infty \rightarrow \infty$, if this numerical renormalization
group equation were exact.  
The deuteron binding energy at $\Lambda_\infty \sim 50$ is off
by more than an MeV.  Table~\ref{table:four} shows 
that 95\% of this difference can indeed be anticipated by examining the
running of the results for $\Lambda_\infty \sim 50$:  
the projected valued at $\Lambda_\infty = 140$  
is correct to $\sim$ 30 keV.  For $b=1.13f$, the case converging
most quickly in Fig. 3, a similar 30 keV accuracy is achieved 
for an extrapolation from still lower $\Lambda_\infty$, 34-40.
Numerically, such an extrapolation can be done quite efficiently:
with proper coding, it costs relatively little ``overhead" to
obtain additional results in the neighborhood of some starting
$\Lambda_\infty$.

\begin{table}
\caption{A ``numerical renormalization group" study of the running
of the deuteron binding energy with $\Lambda_\infty$.  The 
results in the middle column are exact.  Those in the right 
column were obtained from Eq. (21), determined from the variation of the $E_{g.s.}(\Lambda_\infty)$
near $\Lambda_\infty = 50$.
An oscillator parameter is $b=1.7f$, chosen so that this is a
slowly converging case requiring significant extrapolation.
The potential used is Argonne $v18$.}
\label{table:four}
\vspace{0.2cm}
\begin{center}
\begin{tabular}{|c|c|c|}
\hline
\hline
$\Lambda_\infty$ & $E_{g.s.}^{calculated}$ (MeV) & $E_{g.s.}^{projected}$ (MeV) \\
\hline
46 & -0.874 & fit \\ 
50 & -1.069 & fit \\
54 & -1.247 & fit \\
60 & -1.476 & -1.480 \\
70 & -1.771 & -1.773 \\
80 & -1.961 & -1.962 \\
90 & -2.077 & -2.071 \\
100 & -2.143 & -2.128 \\
110 & -2.179 & -2.156 \\
120 & -2.196 & -2.167 \\
130 & -2.204 & -2.172 \\
140 & -2.207 & -2.174 \\ \hline \hline
\end{tabular}
\end{center}
\end{table}

The corresponding exercise was done for ${}^3$He using results
for $\Lambda_\infty$ = 44, 48, and 52.  A very similar 
functional form is found
\begin{equation}
E_{g.s.}(\Lambda_\infty) = -6.906 \mathrm{MeV} + 0.403 \mathrm{MeV} e^{-1.57 (\Lambda_\infty/48)^2}
\end{equation}
While we do not have exact results similar to those in Table~\ref{table:four},
the resulting asymptotic value of -6.906 MeV is in good agreement
with the Faddeev value of 6.895 $\pm$ 0.001 MeV.  Such 10 keV 
accuracy compares well with the Green's function Monte Carlo
result, with its 30 keV accuracy.

Now all of this poses an interesting question: our ``renomalization
group" evolution is a nontrivial one, as it involves a truncation
in harmonic oscillator energies for few-body Slater determinants.
Yet it appears this truncation is leading to a very simple exponential
convergence in results.  This can be compared to effective 
field theory, where the truncation is made on the range of the
potential and the convergence is a weaker 1/$\Lambda_\infty$.
Can one derive and therefore understand this attractive aspect
of a harmonic oscillator mode expansion for bound states?

To attack this problem, we envision setting $\Lambda_\infty$ to
some intermediate scale $\sim$ 50 (around 750 MeV) ---
similar to the numerical examples above --- and performing,
prior to our SM effective theory, a preliminary integration over 
modes above this scale.  If we write
\begin{equation}
H \equiv H_0 + H'~~~~~\mathrm{and}~~~~~H^{eff} = H_0 + H'^{eff}
\end{equation}
where $H_0$ is the harmonic oscillator Hamiltonian and 
$H^{eff}$ is the appropriate effective interaction to use 
below $\Lambda_\infty$, then we find
\begin{equation}
H'^{eff} = H' + H'G^0_{\Lambda_\infty}H'^{eff}
\end{equation}
where $G^0_{\Lambda_\infty}$ is the harmonic oscillator Green's
function corresponding to high-momentum scattering, e.g., 
\begin{eqnarray}
G^0_{\Lambda_\infty}(\vec{r}_1,\vec{r}_2) &=& \sum_{\Lambda > \Lambda_\infty}
{|\Psi^{HO}_\Lambda(\vec{r}_1)\rangle \langle \Psi^{HO}_\Lambda(\vec{r}_2) |
\over E-E_\Lambda} \nonumber \\
&=& G^0(\vec{r}_1,\vec{r}_2) - \sum_{\Lambda \leq \Lambda_\infty}
{|\Psi^{HO}_\Lambda(\vec{r}_1)\rangle \langle \Psi^{HO}_\Lambda(\vec{r}_2) |
\over E-E_\Lambda}
\end{eqnarray}
In Eq. (24) the high-momentum ladder sum involves terms of the form
\begin{equation}
\Psi^{f*}_{SM}(\vec{r}_1) H'(\vec{r}_1) G^0_{\Lambda_\infty}(\vec{r}_1,\vec{r}_2) H'(\vec{r}_2)
\Psi^i_{SM}(\vec{r}_2)
\end{equation}
  
The harmonic oscillator Green's function has the attractive
property that it can be decomposed as
\begin{equation}
        G^0_{\Lambda_\infty} = G^0_{\Lambda_\infty}(r_+,r_-)
\end{equation}
where $r_+ = |\vec{r}_1+\vec{r}_2|$ and $r_- = |\vec{r}_1-\vec{r}_2|$.
Now the restriction to large $\Lambda$ corresponds to short times
and thus to short propagation distances. Thus propagation lengths
should shorten as $\Lambda_\infty$ is increased.  Furthermore
the overall behavior of the Green's function is governed by the
factor $e^{-r_+r_-}$.  Thus the propagator is most contracted 
along the $smaller$ of these two coordinates, $r_< = min(r_+,r_-)$.  As the ladder will
be evaluated between long-wavelength SM states, these observations
suggest expanding $\Psi^{f*}_{SM}(\vec{r}_1) H'(\vec{r}_1)$ and
$H'(\vec{r}_2) \Psi^i_{SM}(\vec{r}_2)$ in Eq. (26) around $r_>$ in
a Taylor series in $r_<$.  (There is a more pedagogical discussion
of this technique and its relevance to effective field theory in
Ref. \cite{lepage}.)  

\begin{figure}[htb!]
\psfig{bbllx=2.5cm,bblly=9.0cm,bburx=17.5cm,bbury=16cm,figure=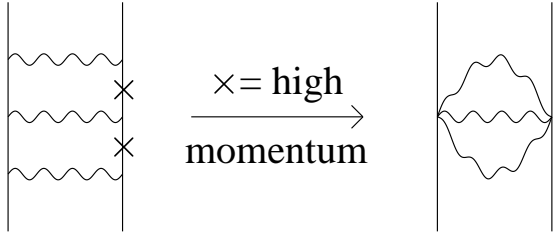,height=2.1in}
\caption{Contraction of a ladder diagram involving only very high
momentum excluded states to a local operator.}
\end{figure}

After a bit of algebra, the net result 
is the contraction of the ladder sum to a local ``rainbow"
diagram, as illustrated in Fig. 6.  The contracted
Green's function has a form familiar to most effective
field theorists,
\begin{equation}
G_{\Lambda_\infty} \rightarrow g^0(r_>) + \vec{\nabla}_{r_>}g^2(r_>)\vec{\nabla}_{r_>}
+ ...
\end{equation}
where this is to be inserted between wave functions evaluated at
$r_>$.  The function $g^0(r_>)$ is obtained
by integrating $G_{\Lambda_\infty}(r_+,r_-)$ over $r_<$,
while $g^2(r_>)$ involves a second moment in the small coordinate
$r_<$, etc.
Effectively this sums the ladder according to a local density
approximation ($g^0$), in an approximation that also takes 
into account the gradient of the local density ($g^2$), etc.

One of the most attractive features of the harmonic oscillator 
is that the full Green's function was recently derived in closed form \cite{macek},
so that $g^0(r_>)$ can be calculated by integrating 
the last expression in Eq. (25):  
an infinite sum over high-momentum modes can be replaced by a
finite sum over low-momentum modes.
The full Green's function is shown in Fig. 7, and the 
integration that produces $g^0$ is depicted in Fig. 8.
The resulting $g^0$ for $\Lambda_\infty$ = 30 and 50 are shown
in Figs. 9 and 10 (dashed lines) along with a similar contraction of the
full Green's function $G^0$.  One finds that the dashed lines --
the residual contribution of very high momentum excitations --
rapidly oscillates, with a frequency very close to $\Lambda_\infty$.
Thus a harmonic oscillator expansion converges not because every
point in coordinate space is well represented by the included
states, but because integration of a product of such oscillations
with a smooth Hamiltonian yields a small remainder.  Note that
increasing $\Lambda_\infty$ also extends the range in $r_>$
spanned by the harmonic oscillator basis functions: extended 
wave functions require more work.

\begin{figure}[htb!]
\psfig{bbllx=-2.0cm,bblly=1.0cm,bburx=15cm,bbury=17.5cm,figure=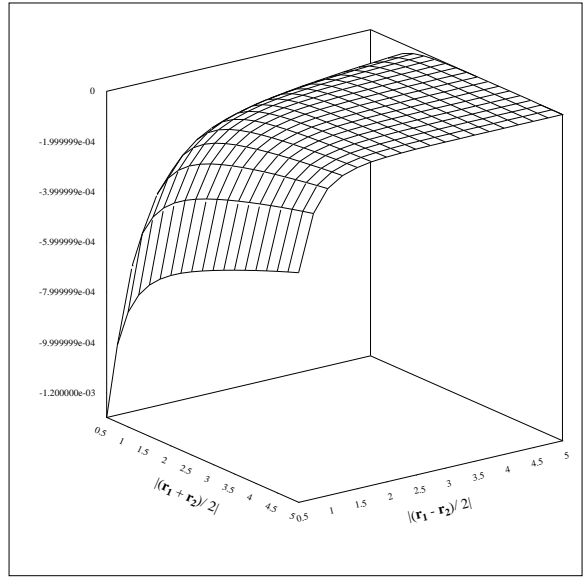,height=3.3in}
\caption{The full (summed over all modes) harmonic oscillator Green's 
function.  For parity conserving interactions, only the even or
odd projections are needed.  The even projection is shown.}
\end{figure}

We are now in the process of implementing this local approximation
to see how far the integration scale $\Lambda_\infty$ can be lowered 
without loss of accuracy.  Unlike effective field theory treatments,
our Taylor series expansion involves a known potential and can
be carried out term by term until exhaustion sets in.  The 
results may provide some estimate of the scales that effective
field theories need to reach to become very accurate.

\begin{figure}[htb!]
\psfig{bbllx=3.0cm,bblly=10.0cm,bburx=14cm,bbury=16cm,figure=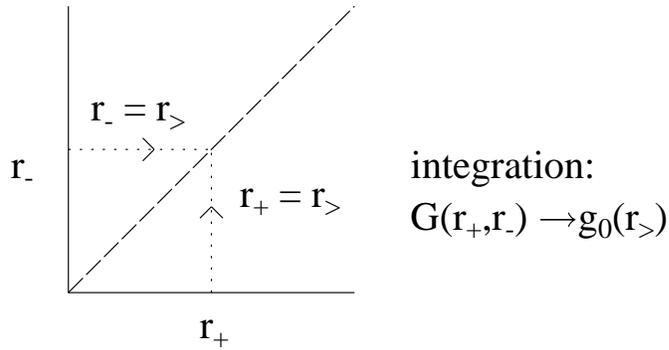,height=1.8in}
\caption{The integration path in the $(r_+,r_-)$ plane that is used
to contract $G(r_+,r_-)$ to $g_0(r_>)$.}
\end{figure}

We will close with two speculations.  If one folds the dashed-line
functions of Figs. 9 and 10 with the potential of Fig. 2, the
result is not dissimilar to a $j_0(\Lambda r_>)$.  Such a
function, integrated with a product of harmonic oscillator 
wave functions, yields a factor 
$e^{-a \Lambda^2}$.
Thus we are hopeful that this line of work may indeed explain
the numerical scaling we have previously discussed.  
Our goal is to add to Table~\ref{table:four} a fourth column in which our
phenomenological scaling function is replaced by a derived one
that yields improved results: the ``break" seen in Table~\ref{table:four}
at $\Lambda_\infty \sim$ 100 (a scale characteristic of the 
Argonne $v18$ hard core) suggests there is a bit of physics
missing from our phenomenological function.  Second,
this suggests that the rapid convergence of the harmonic oscillator
mode expansion, as compared to effective field theory
scaling of 1/$\Lambda_\infty$ which results
from absorbing short-range contributions to the NN potential into contact
terms, may be because it offers a better compromise between
coordinate and momentum space.  Hopefully we will be able to
be more specific soon.

\begin{figure}[htb!]
\psfig{bbllx=-1.0cm,bblly=4.0cm,bburx=17cm,bbury=21.9cm,figure=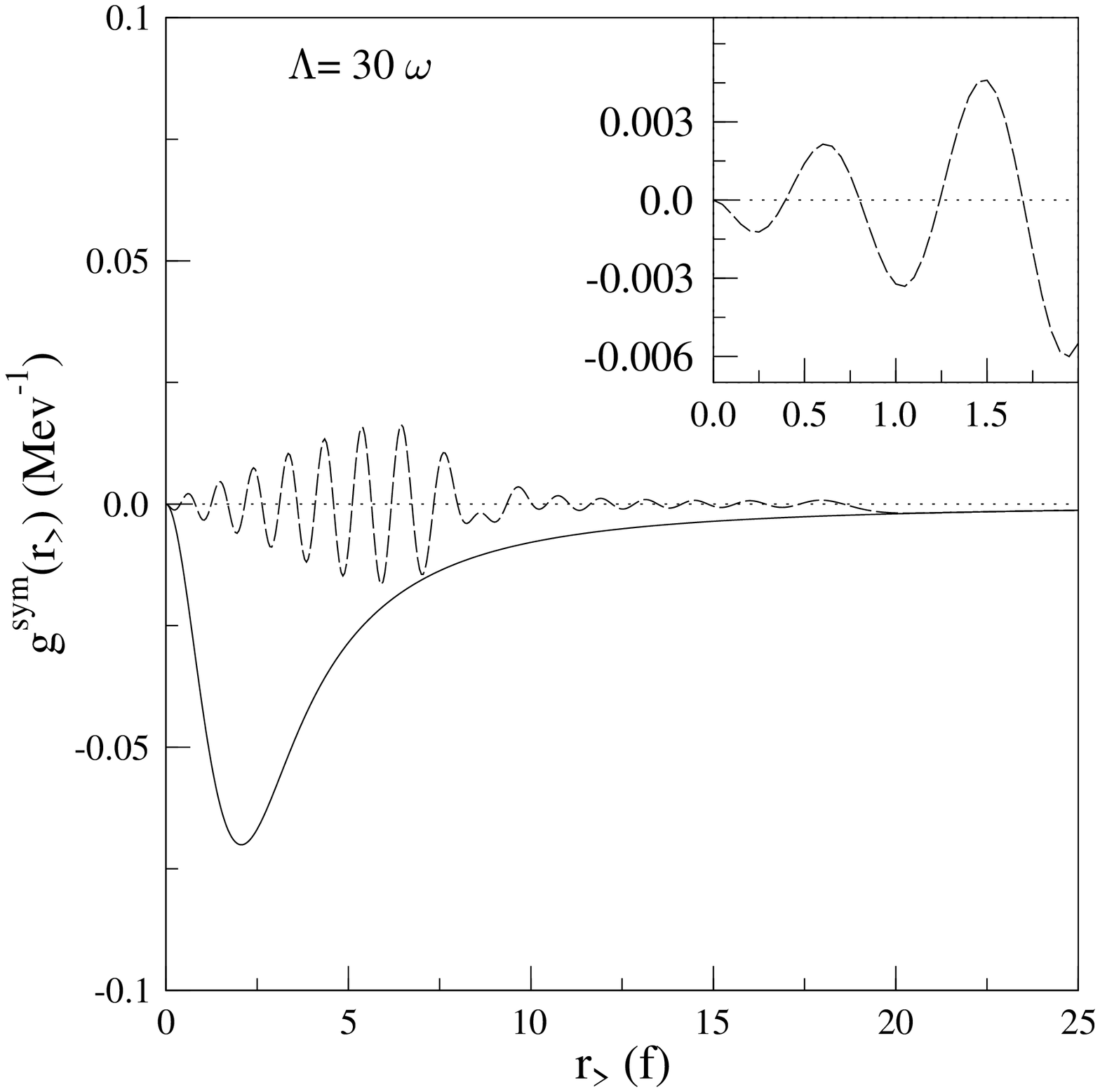,height=3.3in}
\caption{The dashed line is the $g_0(r_>)$ that results when 
excitations $\Lambda > 30$ are summed.  The solid line 
is the corresponding result for the full Green's function.
It is the rapid oscillation of the dashed residual -- rather than 
the accuracy of the harmonic oscillator expansion at all values
of $r_>$ -- that is responsible for the convergence.  The
oscillations are very regular and quite close in frequency 
to $\Lambda \sim 30$.}
\end{figure}

\begin{figure}[htb!]
\psfig{bbllx=-1.0cm,bblly=4.0cm,bburx=17cm,bbury=21.9cm,figure=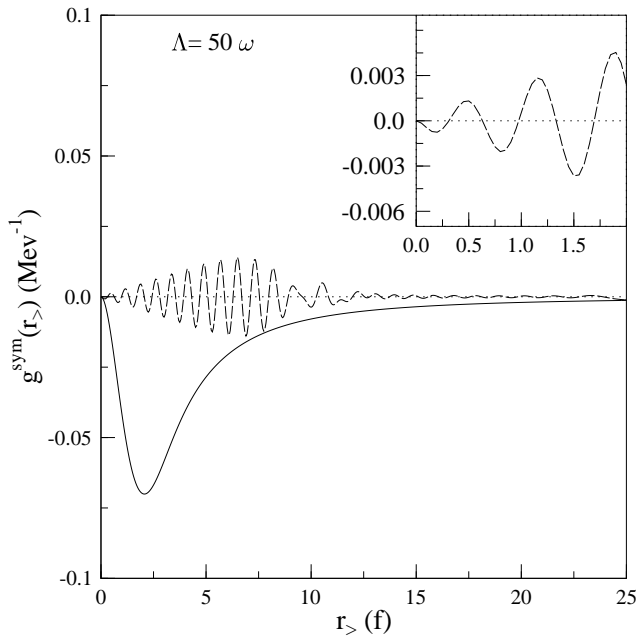,height=3.3in}
\caption{As in Fig. 9, only for $\Lambda > 50$.}
\end{figure}
  
\section{Outlook and Future Plans}

The ability to efficiently solve $^3$He as an effective theory
in a SM-like space is, in itself, quite significant: this
means it is relatively straightforward to execute a faithful
BH treatment of heavier nuclei through order $\rho$ in both
the effective interaction and effective operators.  The
numerical effort is comparable.

The example of $^3$He also suggested that many of the 
uncontrolled approximations made in the SM cannot 
be justified.  Recently sophisticated numerical advances
have been made in SM calculations.  We have argued that it may
be equally important to train this numerical power on the
shakey foundations of that model.  A specific example is our 
use of the Lanczos algorithm -- a favorite among SM practitioners --
to generate the Green's function, leading to a 
tractable strategy for solving the BH equation self-consistently.

We believe that, if the program in Section 3 has modest 
success, we should be able to handle $^4$He with an accuracy
similar to that achieved already for $^3$He.  That would
imply that it is now possible to handle effective interactions
and operators in finite nuclei through order $\rho^2$.
At that point the exciting challenge will be to determine 
where convergence in the cluster expansion is being approached.

We would like to think that the work reported here will 
form a bridge between the ideas growing out of effective
field theories, and the successful phenomenology we have 
achieved in traditional nuclear physics (particularly in 
modeling the NN potential and in solving few-body systems
with nearly exact methods).  It may also stimulate effective
field theory, which has not yet had an impact on nuclear
problems beyond NN and three-body systems.  It is not 
inconceivable that a marriage of EFT -- which strives to handle
few-body problems rigorously, including relativity, 
fundamental few-body forces, etc. --- with some kind of 
cluster expansion of the Brueckner type might someday yield
a controlled theory of finite nuclei.  Furthermore, the
attractive exponential convergence of our harmonic oscillator
mode expansion is also EFT ``food for thought."

This work was supported in part by the Division of Nuclear
Physics, US Department of Energy.

\end{document}